\documentclass[9pt]{article}

\usepackage[a4paper,
            bindingoffset=0.2in,
            left=1.2in,
            right=1.2in,
            top=1in,
            bottom=1in,
            footskip=.25in]{geometry}

\usepackage{amsmath,epsfig,amssymb,amsfonts,amsthm,epsfig,verbatim}
\usepackage{color,graphicx,lscape,longtable}
\usepackage{xcolor,colortbl}
\usepackage{natbib, arydshln}
\usepackage{booktabs}

\def\log{\hbox{log}}

\newcommand{\RED}[1]{{\color{red}{#1}}}
\newcommand{\BLUE}[1]{{\color{blue}{#1}}}

\begin{document}

\title{System Reliability Estimation via Shrinkage}

\author{Beidi Qiang \\
    Department of Math \& Stats, Southern Illinois University Edwardsville\\
    Edsel Pe{\~n}a \\
    Department of Statistics, University of South Carolina}
    \date{}
\maketitle

\begin{abstract}
In a coherent reliability system composed of multiple components configured according to a specific structure function, the distribution of system time to failure, or system lifetime, is often of primary interest. Accurate estimation of system reliability is critical in a wide range of engineering and industrial applications, forming decisions in system design, maintenance planning, and risk assessment. The system lifetime distribution can be estimated directly using the observed system failure times. However, when component-level lifetime data is available, it can yield improved estimates of system reliability. In this work, we demonstrate that under nonparametric assumptions about the component time-to-failure distributions, traditional estimators such as the Product-Limit Estimator (PLE) can be further improved under specific loss functions. We propose a novel methodology that enhances the nonparametric system reliability estimation through a shrinkage transformation applied to component-level estimators. This shrinkage approach leads to improved efficiency in estimating system reliability.
\end{abstract}

{\bf keywords:} Coherent system, simultaneous estimation, shrinkage estimator, nonparametric methods


\section{Introduction}\label{sec:Intro}

Reliability engineering plays an essential role in ensuring the safe, efficient, and sustained operation of complex systems across a wide range of sectors, including manufacturing, aerospace, healthcare, and software development. The ability to accurately assess and predict the reliability of the system is fundamental to robust design, proactive maintenance scheduling, and comprehensive risk management. As systems become increasingly intricate, comprising numerous interconnected components, the challenge of precisely estimating their overall reliability becomes more pronounced.

The lifetime distribution of a coherent system can be estimated directly from observations of system-level failure times. However, when component-level data are available, more accurate estimation can be achieved by first estimating the component-level reliability functions and then combining them, according to the system structure function, to estimate system reliability. The literature on system reliability estimation has explored a variety of approaches for different data collection schemes, including cases where only system-level data are available and those in which detailed component-level (or ``autopsy”) data are collected. Yet a key question remains largely unaddressed: can these standard nonparametric estimators be systematically improved, particularly under a decision-theoretic framework that explicitly incorporates a specific loss function?

This research addresses this important gap by proposing and evaluating a novel class of shrinkage estimators for nonparametric system reliability functions. We recognize that the problem of estimating component-level reliabilities is fundamentally one of simultaneous estimation. Motivated by this, we adopt the idea of the James–Stein shrinkage estimator \cite{Stein1961}, which is well known to dominate the standard least squares estimator under quadratic loss when simultaneously estimating the means of independent Gaussian variables. This idea naturally extends to the reliability setting, specifically, to the problem of estimating system reliability from component-level lifetime data. In our proposed approach, shrinkage-type estimators are first obtained for each component's reliability function. These are then combined, using the known structure function of the system, to yield an improved estimator of the overall system reliability under an invariant global loss function.

Shrinkage estimation has previously been explored in reliability analysis. Past work has proposed shrinkage estimators of component reliabilities under various lifetime models and extended the ideas to system-level estimation for certain structures and parametric assumptions; see, for example, \cite{Pandey1987}. The common theme in such approaches is the application of a shrinkage transformation to maximum likelihood estimators to reduce their risk under a global loss. In our earlier work \cite{Qiang2018}, we proposed an improved estimator of system reliability based on parametric lifetime assumptions. A key strength of that approach is its generality: rather than shrinking parameter estimates, it targets the estimated component hazard functions, making it adaptable to more flexible modeling frameworks.

In this paper, we extend that approach to a fully nonparametric setting. This extension removes the need for strong distributional assumptions and enhances the flexibility of the methodology, allowing it to better accommodate a wide range of failure behaviors encountered in practice. The remainder of the paper is organized as follows. Section~\ref{sec:background} reviews an essential background on coherent systems and their associated structure and reliability functions. Section~\ref{sec:framework} introduces the data settings considered and formulates the nonparametric decision-theoretic framework. Section~\ref{sec:improvedEstimation} presents the proposed estimators, along with analytical and bootstrap methods for determining the optimal shrinkage coefficient. Section~\ref{sec:simulation} reports on simulation studies and discusses the results, implications, and limitations. Finally, Section~\ref{sec:Conclusion} concludes with a summary of contributions and directions for future work.


\section{Background on Reliability Systems and Functions} \label{sec:background}

This section establishes the foundational concepts of reliability systems and their associated functions. For general background and definitions in reliability theory, we refer the reader to the classic text by \cite{BarPro81}.

Two distinct data accrual schemes are considered, which determine the level of information available about system and component lifetimes. In the first scheme, only system-level data are collected—specifically, observations of the system lifetimes. This type of data collection is often straightforward but provides limited insight into the contributions of individual components to system failure. A more informative scheme involves the observation of lifetimes for individual components within each system. In this setting, component lifetimes may be fully observed or right-censored, depending on the system failure time or an upper bound on the monitoring period. This richer data enables more granular analysis of the impact of each component on overall system performance.

A pervasive challenge in lifetime data analysis, particularly in reliability studies, is the presence of censoring. Right censoring occurs when the exact failure time of a system or component is unknown, but it is known to have survived up to a certain point in time. Common causes of right censoring include:

\begin{itemize}
    \item \textbf{Finite monitoring periods:} Studies often have a predefined endpoint, beyond which no additional information is collected. Units that have not failed by this time are right-censored.
    \item \textbf{System withdrawal from observation:} A system may be removed from observation for reasons unrelated to failure, such as scheduled upgrades or external incidents.
    \item \textbf{Component censoring by system failure:} In multi-component systems, components that are still functioning when the system fails are right-censored at the system’s failure time.
\end{itemize}

The presence of censoring complicates statistical inference, as it introduces incomplete data and necessitates specialized estimation techniques that can handle unobserved failure times.

\subsection{Coherent Reliability Systems} \label{sec:system}

Consider a system consisting of \(K\) independent components with a lifetime variable \(S\). The structure of a coherent system is defined by a structure function \( \phi: \{0,1\}^K \to \{0,1\} \). Each component \(j\) has a state variable \(X_j(t) = I\{T_j > t\}\), indicating whether it is operational at time \(t\), where \(T_j\) denotes the lifetime of component \(j\). The overall system state at time \(t\) is given by \(\phi(X_1(t),\ldots,X_K(t))\), and the system lifetime is defined as:
\[
S = \sup\{t \geq 0 : \phi[X_1(t),\ldots,X_K(t)] = 1\}.
\]
Accordingly, the system reliability function is
\[
R_S(t) = \Pr\{S > t\} = \mathbb{E}[\phi(X_1(t),\ldots,X_K(t))] = h_\phi[R_1(t),\ldots,R_K(t)],
\]
where \(R_j(t) = \Pr\{T_j > t\}\) is the reliability function of component \(j\), and \( h_\phi(\cdot) \) is determined by the system’s configuration.

\subsection{Examples of System Structures} \label{sec:egOfstructure}

We now illustrate the structure function framework using several canonical system configurations:

\begin{itemize}
    \item \textbf{Series system:} The system operates only if all components are operational. It fails upon the failure of any single component. The structure and reliability functions are:
    \[
    \phi_{ser}(x_1,\ldots,x_K) = \prod_{i=1}^K x_i, \quad R_{ser}(t) = \prod_{i=1}^K R_i(t).
    \]

    \item \textbf{Parallel system:} The system functions as long as at least one component is operational. It fails only when all components fail. The corresponding structure and reliability functions are:
    \[
    \phi_{par}(x_1,\ldots,x_K) = 1 - \prod_{i=1}^K (1 - x_i), \quad R_{par}(t) = 1 - \prod_{i=1}^K (1 - R_i(t)).
    \]

    \item \textbf{Series-parallel system:} These configurations combine series and parallel subsystems. For example, in a 3-component system where component 1 is in series with a parallel subsystem comprising components 2 and 3, the structure and reliability functions are:
    \[
    \phi_{serpar}(x_1,x_2,x_3) = x_1[1-(1-x_2)(1-x_3)], 
      \] 
    \[
    R_{serpar}(t) = R_1(t)[1 - (1 - R_2(t))(1 - R_3(t))].
    \]
\end{itemize}

\subsection{Component Importance} \label{sec:importance}

A key objective in system analysis is identifying the contribution of each component to overall system reliability. A widely used measure is the \textit{reliability importance} of component \(j\), defined as the partial derivative:
\begin{equation}
I_\phi(j;\mathbf{p}) = \frac{\partial h_\phi(p_1,\ldots,p_K)}{\partial p_j} = h_\phi(\mathbf{p},1_j) - h_\phi(\mathbf{p},0_j),
\end{equation}
where \(\mathbf{p} = (p_1,\ldots,p_K)\) is the vector of component reliabilities, and \((\mathbf{p},1_j)\) denotes replacing \(p_j\) with 1. For coherent systems, it holds that \(I_\phi(j;\mathbf{p}) > 0\). This measure quantifies the sensitivity of system reliability to each component.

For the system types introduced in Section~\ref{sec:egOfstructure}, the importance functions simplify as follows:

\begin{itemize}
    \item \textbf{Series system:} \(I_{ser}(j; \mathbf{p}) = {h_{ser}(\mathbf{p})}/{p_j}\). This indicates that the least reliable component is the most critical, coinciding with the saying that a {\em system is as good as its weakest component.}
    \item \textbf{Parallel system:} \(I_{par}(j; \mathbf{p}) = {[1 - h_{par}(\mathbf{p})]}/{[1 - p_j]}\). This indicates that the most reliable component is the most influential.
    \item \textbf{Series-parallel system} (3 components): 
    \[
    I_{serpar}(1; \mathbf{p}) = 1 - (1 - p_2)(1 - p_3), \quad 
    I_{serpar}(2; \mathbf{p}) = p_1(1 - p_3), \quad 
    I_{serpar}(3; \mathbf{p}) = p_1(1 - p_2).
    \]
    For example, when all components have equal reliability \(p\), we find \(I_{serpar}(1; p) = p(2 - p)\) and \(I_{serpar}(2; p) = I_{serpar}(3; p) = p(1 - p)\), confirming that the series component (component 1) is the most influential. However, if component 1 is highly reliable while components 2 and 3 are not, the relative importance may shift.
\end{itemize}

\subsection{Nonparametric Estimation of System Reliability} \label{sec:npsys}

We now turn to the estimation of the system reliability function based on observed data, either at the system or component level.

\paragraph{System-level data.}

When only system lifetimes are observed, a natural nonparametric estimator of \(R_S(t)\) is the Kaplan-Meier (KM) or product-limit estimator \cite{KapMei58, KalPre80, FleHar91, ABGK93}. Suppose we observe system data of the form:
\[
\mathbf{D} = \left\{(Z_{i},\delta_{i}): i=1,2,\ldots,n\right\},
\]
where \(Z_i\) is the observed (possibly censored) system lifetime and \(\delta_i\) is the failure indicator.

In this setting, component-level information is unavailable, and parameter identifiability can be problematic under parametric assumptions. A nonparametric approach avoids this issue by estimating \(R_S(t)\) without assuming a specific distributional form. The product-limit estimator is given by:
\begin{equation}
\label{eq: ple}
\hat R_S(t)=\prod_{\{m:\ t_{(m)}\le t\}}\left[ 1-\frac{d_m}{n_m}\right],
\end{equation}
where \(t_{(1)} < \cdots < t_{(M)}\) are the ordered distinct uncensored times, \(d_m\) is the number of failures at \(t_{(m)}\), and \(n_m\) is the number at risk just prior to \(t_{(m)}\). That is,
\[
d_m = \sum_{i=1}^n I\{Z_i = t_{(m)}, \delta_i = 1\}, \quad
n_m = \sum_{i=1}^n I\{Z_i \ge t_{(m)}\}.
\]

Under mild conditions, \(\hat{R}_S(t)\) is asymptotically normal as \(n \rightarrow \infty\), with asymptotic variance estimated via Greenwood’s formula:
\begin{equation}
\label{eq:greenwood}
\widehat{\hbox{Avar}[\sqrt{n} \hat{R}_S(t)]} =
[\hat{R}_S(t)]^2 \sum_{\{l: t_{(l)} \le t\}} \frac{d_l}{n_l (n_l - d_l)}.
\end{equation}

\paragraph{Component-level data.}

When component lifetimes are observed or censored, we can leverage this information and the known system structure \(\phi(\cdot)\) to obtain an improved system reliability estimator. Let the observed data be:
\begin{equation}
\label{observable data}
\mathbf{D} = \left\{(Z_{ij},\delta_{ij}): i=1,2,\ldots,n; j=1,2,\ldots,K\right\},
\end{equation}
where each \(Z_{ij}\) is subject to right-censoring by both the system lifetime \(S_i\) and a monitoring time \(C_i\):
\[
Z_{ij} = \min\{T_{ij}, \min(C_i, S_i)\}, \quad \delta_{ij} = I\{T_{ij} \le \min(C_i, S_i)\}.
\]

The system reliability function can be expressed via the structure function:
\begin{equation}
R_S(t) = h_\phi[R_1(t), \ldots, R_K(t)].
\label{system reliability function}
\end{equation}
Each component reliability function \(R_j(t)\) can be estimated using the KM estimator \(\hat R_j(t)\). The plug-in estimator for \(R_S(t)\) is then:
\[
\hat R_S(t) = h_\phi[\hat R_1(t), \ldots, \hat R_K(t)].
\]

The important observation to note is that the estimation of the component reliabilities \(R_j(t)\) constitutes a simultaneous estimation problem. This observation motivates the application of shrinkage techniques, which can improve estimation efficiency under appropriate loss functions. The asymptotic properties of each \(\hat R_j(t)\) mirrors those of \(\hat R_S(t)\), including its asymptotic normality and the use of Greenwood's variance estimate.

\section{Decision-Theoretic Framework and Shrinkage Estimators}\label{sec:framework}

This section introduces the decision-theoretic framework used to develop and evaluate the proposed shrinkage estimators. This principled framework assesses estimator performance based on expected loss under a specified criterion and provides a rigorous foundation for comparing and improving estimators.

\subsection{Decision-theoretic Framework}\label{sec:dtFramework}
We begin with a parametric decision-theoretic formulation, which we later extend to nonparametric settings through the choice of an appropriate loss function. A classical statistical decision problem is characterized by the following elements:
\[
(\Theta, \mathcal{A}, \mathcal{X}, \mathcal{F} = \{F(\cdot|\boldsymbol{\theta}), \boldsymbol{\theta} \in \Theta\}, L, \mathcal{D}).
\]
For notation, given a space $\mathcal{S}$, $\mathfrak{S} \equiv \sigma(\mathcal{S})$ will denote an appropriate sigma-field of subsets of $\mathcal{S}$, e.g., if $\mathcal{S}$ is a metric space, then $\mathfrak{S}$ will be the Borel $\sigma$-field of $\mathcal{S}$ determined by the metric. Here, \(\Theta\) denotes the parameter space, which may be finite- or infinite-dimensional, and contains all possible values of the unknown parameter vector \( \boldsymbol{\theta} \). The action space \(\mathcal{A}\) consists of all possible decisions the decision-maker can make. The value of the measurable\footnote{with respect to the product $\sigma$-field $\mathfrak{\Theta} \otimes \mathfrak{A}$ on $\Theta \times \mathcal{A}$ and the Borel $\sigma$-field $\mathfrak{R}$ on $\mathbb{R}$.} loss function \(L: \Theta \times \mathcal{A} \rightarrow \mathbb{R}\), $L(\boldsymbol{\theta},\mathbf{a})$, quantifies the loss or penalty incurred by taking action \(\mathbf{a}\) when the true parameter value is \(\boldsymbol{\theta}\).

The observable (data) random variable \(X\) takes values in the sample space \(\mathcal{X}\), and its distribution, conditional on \(\boldsymbol{\theta}\), is denoted by \(F(\cdot|\boldsymbol{\theta}) \in \mathcal{F}\). A (non-randomized) decision rule or estimator is a measurable function \(\delta: \mathcal{X} \rightarrow \mathcal{A}\), and the set of all such functions is represented by \(\mathcal{D}\).

To evaluate a decision rule \(\delta \in \mathcal{D}\), we use the risk function:

\[
\text{Risk}(\boldsymbol{\theta}, \delta) = \mathbb{E}[L(\boldsymbol{\theta}, \delta(X)) \mid \boldsymbol{\theta}],
\]
which gives the expected loss incurred by \(\delta\) when the true parameter is \(\boldsymbol{\theta}\). A rule \(\delta_1\) is said to dominate another rule \(\delta_2\) if:
\[
\text{Risk}(\boldsymbol{\theta}, \delta_1) \leq \text{Risk}(\boldsymbol{\theta}, \delta_2) \quad \text{for all } \boldsymbol{\theta} \in \Theta,
\]
with strict inequality for at least one value of \(\boldsymbol{\theta}\). In such cases, \(\delta_2\) is considered inadmissible.


\subsection{James-Stein Shrinkage Estimators}\label{sec:js}

This framework naturally extends to simultaneous estimation problems. Consider the case where
\(\boldsymbol{\theta} = (\mu_1, \mu_2, \ldots, \mu_K) \in \mathbb{R}^K\), and we observe \(\mathbf{X} = (X_1, \ldots, X_K) \sim \mathcal{N}(\boldsymbol{\mu}, \sigma^2 I_K)\), where the components are independent and \(\sigma^2\) is {\em known}. Under the squared-error or quadratic loss function
\[
L(\boldsymbol{\theta}, \mathbf{a}) = \|\boldsymbol{\theta} - \mathbf{a}\|^2 = \sum_{i=1}^K (\mu_i - a_i)^2,
\]
the maximum likelihood (ML) estimator given by \(\delta_{ML}(\mathbf{X}) = \mathbf{X}\) has the constant risk
\[
\text{Risk}(\boldsymbol{\theta}, \delta_{ML}) = K \sigma^2.
\]

While this ML estimator is admissible for \(K = 1\) or \(K = 2\), it is no longer admissible when \(K \geq 3\). In this case, James and Stein \cite{Stein1961} showed that shrinkage estimators can outperform \(\delta_{ML}\) under the quadratic loss. The James-Stein estimator is given by:
\[
\delta_{JS}(\mathbf{X}) = \left[1 - \frac{(K - 2)\sigma^2}{\|\mathbf{X}\|^2}\right] \mathbf{X},
\]
which shrinks the ML estimate toward the origin. This shrinkage reduces risk by introducing bias in exchange for reduced variance.

More generally, suppose \(\mathbf{X}_1, \ldots, \mathbf{X}_n\) are i.i.d. \(K\)-dimensional normal vectors with mean \(\boldsymbol{\theta}\) and covariance \(\sigma^2 I_K\), again with $\sigma$ known. Then the James-Stein estimator based on the sample mean \(\overline{\mathbf{X}} = \frac{1}{n} \sum_{i=1}^n \mathbf{X}_i\) becomes:
\[
\delta_{JS}(\overline{\mathbf{X}}) = \left[1 - \frac{(K - 2)\sigma^2}{n \|\overline{\mathbf{X}}\|^2}\right] \overline{\mathbf{X}}.
\]

This estimator demonstrates the key advantage of shrinkage: even for independent variables, borrowing information across coordinates leads to improved estimation under a global loss criterion.

While the James-Stein estimator was originally developed for multivariate normal settings, the idea has since been generalized to broader contexts. In particular, it has been adapted to nonparametric problems of location estimation; see \cite{Qiang2023} for details.

Shrinkage estimators typically take the form:
\[
\delta_c(\mathbf{X}) = c \, \overline{\mathbf{X}},
\]
where \(c > 0\) is a data-dependent multiplier. Note that `shrinkage' usually refers to the situation where $c \in (0,1]$, but it is possible in our development for $c > 1$, and even in this case we shall still use the adjective `shrinkage'. This idea underlies the development of improved estimators in our reliability setting.


\subsection{Shrinkage Estimation in Reliability Analysis}\label{sec:ShrinkagInRel}

Traditionally, system reliability is estimated by first obtaining component-level MLEs of the components reliabilities and then plugging-in these estimates in the system reliability function to obtain an estimate of the system-level reliability. The concept of simultaneous estimation via shrinkage extends naturally when estimating system reliability from component-level data. Shrinkage estimators for reliability have been studied under parametric lifetime distributions, including exponential and Weibull models. For example, Siu and Keung \cite{SiuKeung1996} considered shrinkage estimation for exponential lifetimes, and Prakash et al. \cite{Prakash2008} extended these ideas using the LINEX loss under censoring. Further works, including \cite{Pandey1985, Pandey1987, Pandey2014}, explored shrinkage estimation for components and systems under various lifetime models. In our previous work \cite{Qiang2018}, we investigated parametric shrinkage for system reliability, showing that improved estimation could be achieved via component-level hazard function shrinkage. A distinctive feature of our approach is that shrinkage is applied not to the distributional parameters, but directly to the estimated hazard or reliability functions. This allows our method to be extended naturally to nonparametric settings.

\paragraph{Loss Function and Risk.} To adapt the shrinkage approach to nonparametric estimation, we adopt the weighted Cramér–von Mises loss:
\[
L(R_S(\cdot), \hat{R}_S(\cdot)) = - \int \frac{[R_S(t) - \hat{R}_S(t)]^2}{R_S(t)(1 - R_S(t))} \, dR_S(t).
\]
%
%
This loss function is a global measure of a standardized measure of the difference between the estimator of the system reliability function and the true system reliability function.
%
The expected loss (or risk) associated with an estimator \(\hat{R}_S(\cdot)\) is:
\[
\text{Risk}(R_S(\cdot), \hat{R}_S(\cdot)) = \mathbb{E} \left[ - \int \frac{[R_S(t) - \hat{R}_S(t)]^2}{R_S(t)(1 - R_S(t))} \, dR_S(t) \right] =\mathbb{E} \left[ \int \frac{[R_S(t) - \hat{R}_S(t)]^2}{R_S(t)(1 - R_S(t))} \, dF_S(t) \right],
\]
where the expectation is taken over repeated samples under the data-generating process. The goal is to construct estimators that minimize this risk. This forms the foundation of our proposed shrinkage estimators for nonparametric system reliability estimation.


\section{Shrinkage Estimation with Autopsy Data}
\label{sec:improvedEstimation}

In the classical approach, system reliability is estimated using product-limit (PL) estimators for each component's reliability function. The plug-in method for component-level data involves the simultaneous estimation of multiple functions \( R_j(t) \). Shrinkage techniques, inspired by the James-Stein estimator, can be applied to improve the overall efficiency of the system reliability estimator. By leveraging structural and importance-based relationships among the components, shrinkage approaches can significantly enhance finite-sample performance, especially when the sample size is small and the number of components \( K \) is large. This section introduces an improved system reliability estimation procedure using shrinkage techniques and explores both analytical and computational methods for determining an optimal shrinkage coefficient \( c^* \).

\subsection{The Class of Shrinkage Estimators}
\label{subsec-reliability-estimation}

Let \( R_1(t), \dots, R_K(t) \) denote the reliability functions of the \( K \) components. The system reliability function \( R_S(t) \) is determined by the structure function \( h_\phi \), which combines the component reliabilities:
\[
R_S(t) = h_\phi(R_1(t), \dots, R_K(t)).
\]

We consider a class of shrinkage estimators where a shrinkage transformation is applied to each component's PL estimator. Specifically, for a given coefficient \( c \), the component estimator becomes \( \hat{R}_j(t)^c \), and the system reliability estimator is defined as:
\[
\hat{R}_S(t;c) = h_\phi(\hat{R}_1(t)^c, \dots, \hat{R}_K(t)^c).
\]
Note that when \( c = 1 \), the estimator \( \hat{R}_S(t;1) \) corresponds to the standard plug-in estimator.

Furthermore, if \( \hat{\Lambda}_j(t) \) denotes an estimator of the cumulative hazard function \( \Lambda_j(t) \) for component \( j \), then we have the relation:
\[
[\hat{R}_j(t)]^c = \exp[-c \hat{\Lambda}_j(t)],
\]
illustrating that applying shrinkage to the PL estimators is equivalent to applying shrinkage to the cumulative hazard function estimators. This formulation does not require direct shrinkage of any parameter associated with a specific distribution, making the method flexible and broadly applicable in nonparametric settings.

To operationalize this estimator, we must determine the optimal shrinkage coefficient \( c \) using the observed data. The following subsections describe two approaches for this: an analytical method and a bootstrap-based method.

\subsection{Analytical Approach to Determining Optimal \( c^* \)}
\label{sec: AnalyticalC}
The risk function associated with a given shrinkage coefficient \( c \) is defined as:
\begin{align}
\text{Risk}(c) &= \mathbb{E} \left[ - \int \frac{[\hat{R}_S(t;c)-R_S(t)]^2}{R_S(t)(1-R_S(t))} \, dR_S(t) \right] \\
&= \mathbb{E} \left[ \int \frac{[h_\phi\{[\hat{R}_1(t)]^c, \dots, [\hat{R}_K(t)]^c\} - R_S(t)]^2}{R_S(t)(1-R_S(t))} \, dF_S(t) \right].
\end{align}

At the extremes, as \( c \to 0 \), the estimator converges to \(\hat{R}_S(t;c) \to h_\phi\{1, \dots, 1\} = 1\), while as \( c \to \infty \), \(\hat{R}_S(t;c) \to h_\phi\{0, \dots, 0\} = 0\). In both cases, the risk diverges to infinity. Since the risk function is continuous in \( c \), an optimal value \( c^* \in (0, \infty) \) must exist.

To approximate \( \text{Risk}(c) \), we perform a first-order Taylor expansion of \( \hat{R}_S(t;c) \) about \( R_j(t) \):
\[
\hat{R}_S(t;c) \approx h_\phi(R_1(t), \dots, R_K(t)) + \sum_{j=1}^K I_\phi(j,t) \,[\hat{R}_j(t)^c - R_j(t)],
\]
where \( I_\phi(j,t) \) is the reliability importance of component \( j \) at time \( t \). Substituting into the risk function gives:
\begin{align*}
R(R_S, \hat{R}_S(t;c)) &\approx \mathbb{E} \left[\int \frac{\left\{ \sum_{j=1}^K I_\phi(j,t) [\hat{R}_j(t)^c - R_j(t)] \right\}^2}{R_S(t)(1 - R_S(t))} \, dF_S(t) \right] \\
&= \int \sum_{j=1}^K \frac{I_\phi(j,t)^2 \, \mathbb{E}[(\hat{R}_j(t)^c - R_j(t))^2]}{R_S(t)(1 - R_S(t))} \, dF_S(t) \\
&\quad + \int \sum_{j \neq k} \frac{I_\phi(j,t) I_\phi(k,t) \, \mathbb{E}[(\hat{R}_j(t)^c - R_j(t))(\hat{R}_k(t)^c - R_k(t))]}{R_S(t)(1 - R_S(t))} \, dF_S(t).
\end{align*}

Using the Delta method and asymptotic normality of the product-limit estimators:
\[
\text{Var}[\hat{R}_j(t)^c] \approx c^2 R_j(t)^{2c - 2} V_j,
\]
which implies:
\[
\mathbb{E}[(\hat{R}_j(t)^c - R_j(t))^2] \approx c^2 R_j(t)^{2c - 2} V_j + [R_j(t)^c - R_j(t)]^2.
\]
Assuming independence between components, simplifies the cross terms to:
\[
\mathbb{E}[(\hat{R}_j(t)^c - R_j(t))(\hat{R}_k(t)^c - R_k(t))] \approx [R_j(t)^c - R_j(t)][R_k(t)^c - R_k(t)].
\]
The approximation, \( E[\hat{R}_j^c(t)] \approx R_j^c(t) \), is based on the invariant and asymptotic property of PLE, which is also the maximum likelihood estimation of the discrete hazard function for each component.

The approximate risk thus becomes:
\begin{align*}
\text{Risk}(c) &\approx \int \sum_{j=1}^K \frac{I_\phi(j,t)^2 [c^2 R_j(t)^{2c - 2} V_j + (R_j(t)^c - R_j(t))^2]}{R_S(t)(1 - R_S(t))} \, dF_S(t) \\
&\quad + \int \sum_{j \neq k} \frac{I_\phi(j,t) I_\phi(k,t) (R_j(t)^c - R_j(t))(R_k(t)^c - R_k(t))}{R_S(t)(1 - R_S(t))} \, dF_S(t).
\end{align*}

In practice, \( R_j(t) \) and \( V_j \) are replaced by their empirical counterparts \(\hat{R}_j(t)\) and \(\hat{V}_j(t)\) (with \(\hat{V}_j(t)\) estimated using Greenwood’s formula; see \eqref{eq:greenwood}). The empirical risk is minimized numerically (e.g., via Nelder–Mead \cite{Nelder1965} or quasi-Newton \cite{Broyden1967}) to yield the estimated optimal \( \hat{c}^* \). The shrinkage-enhanced estimator is then:
\[
\hat{\hat{R}}_S(t) = h_\phi[\hat{R}_1(t)^{\hat{c}^*}, \ldots, \hat{R}_K(t)^{\hat{c}^*}].
\]
As \( n \to \infty \), \( V_j \to 0 \) and the risk simplifies to:
\[
\text{Risk}(c) \approx \int \sum_{i,j} \frac{I_\phi(i,t) I_\phi(j,t) (R_i(t)^c - R_i(t))(R_j(t)^c - R_j(t))}{R_S(t)(1 - R_S(t))} \, dF_S(t),
\]
which is minimized at \( c = 1 \), confirming that shrinkage primarily benefits small-sample or high-variance settings.

\medskip
\noindent\textbf{Sign of \( c^* \) and system structure.}  
The value of \( c^* \) is not necessarily below 1—it depends on system structure. Expanding around \( c = 1 \) gives:
\[
R_j^c(t) \approx R_j(t)+(c-1)R_j(t)\log R_j(t),
\]
\[
R_j^{2c}(t) \approx [R_j(t)]^2+2(c-1)[R_j(t)]^2\log R_j(t),
\]
leading to the quadratic approximation:
\begin{equation}
\label{polynomial risk}
\text{Risk}(c) \approx A\cdot c^2 + 2B\cdot c^2(c-1) + D\cdot (c-1)^2,
\end{equation}
where:
\begin{eqnarray*}
A&=& \int_0^\infty \sum_{j=1}^K \frac {I^2_\phi(j,t) V_j(t)}{R(t)[1-R(t)]}dF(t) > 0, \\
B&=& \int_0^\infty \sum_{j=1}^K \frac {I^2_\phi(j,t)V_j(t)\log R_j(t) }{R(t)[1-R(t)]}dF(t) < 0, \\
D&=&\int_0^\infty \frac {[\sum_{j=1}^K I_\phi(j,t)R_j(t)\log R_j(t)]^2}{R(t)[1-R(t)]}dF(t) > 0.
\end{eqnarray*}
The minimizer of \eqref{polynomial risk} is:
\[
c^*[R_1(t),\dots,R_K(t)] = \frac{2B - A - D + \sqrt{(2B - A - D)^2 + 12BD}}{6B}.
\]
The derivative at \( c = 1 \) is \( \text{Risk}'(1) \approx 2(A+B) \). If \( A+B > 0 \), or equivalently   
\begin{equation}
\label{AplusB}
\int_0^\infty \sum_{j=1}^K \frac {I^2_\phi(j,t) V_j(t)[1+\log R_j(t)]}{R(t)[1-R(t)]}dR(t)<0,
\end{equation}
the approximated risk is increasing at \(c=1\), which means the optimal \(c^*\) is below 1. If \( A+B < 0 \), then \( c^* > 1 \). In this framework, \( c^* > 1 \) corresponds to shrinking component reliabilities downward, while \( c^* < 1 \) corresponds to inflating them.

For a \emph{series system}:
\[
I_\phi(j,t) = \frac{R(t)}{R_j(t)},
\]
so condition \eqref{AplusB} becomes:
\[
\int_0^\infty  \frac{R(t)}{1-R(t)} \sum_{j=1}^K \frac {1+\log R_j(t)}{R_j^2(t)}  V_j(t)\,dR(t).
\]
Here, the term \([1+\log R_j(t)]/R_j^2(t)\) takes larger absolute values when negative, favoring \( A+B > 0 \) and hence \( c^* < 1 \).

For a \emph{parallel system}:
\[
I_\phi(j,t) = \frac{1-R(t)}{1-R_j(t)},
\]
leading to:
\[
\int_0^\infty  \frac{1-R(t)}{R(t)} \sum_{j=1}^K \frac {1+\log R_j(t)}{[1-R_j(t)]^2}  V_j(t)\,dR(t),
\]
where the corresponding term is typically positive, favoring \( c^* > 1 \).

These analytical insights align with simulation results: \( c^* \) tends to be below 1 for series systems and above 1 for parallel systems. Intuitively, shrinkage here serves to moderate the relative importance of components—dampening the effect of overly dominant or underperforming components and pulling importance values toward a balanced middle ground.

\subsection{Bootstrap Approach to Determining Optimal \( c^* \)}
\label{sec: BootstrapC}

An alternative method to determine \( c^* \) is through the bootstrap. This approach avoids asymptotic approximations and is data-driven:

\begin{itemize}
    \item For \( b = 1, \dots, B \), resample with replacement from the observed dataset \( \mathbf{D} = \{(Z_{ij}, \delta_{ij})\} \) to obtain \( \mathbf{D}_b^* \).
    \item For each bootstrap sample \( \mathbf{D}_b^* \), compute the PL estimators \( \hat{R}_j(t; \mathbf{D}_b^*) \), for \( j = 1, \dots, K \).
    \item For each \( c \in \{c_1, \dots, c_M\} \subset \mathbb{R}_+ \), compute the plug-in shrinkage estimator:
    \[
    \hat{R}_S(t; \mathbf{D}_b^*, c) = h_\phi[\hat{R}_1(t; \mathbf{D}_b^*)^c, \ldots, \hat{R}_K(t; \mathbf{D}_b^*)^c].
    \]
    \item Compute the loss for each bootstrap sample:
    \[
    L_b(c) = L(\hat{R}_S(t; \mathbf{D}), \hat{R}_S(t; \mathbf{D}_b^*, c)).
    \]
    \item Average the loss over all \( B \) samples:
    \[
    \widehat{\text{Risk}}(c) = \frac{1}{B} \sum_{b=1}^B L_b(c).
    \]
    \item Choose \( \hat{c}^* \) as the minimizer of \( \widehat{\text{Risk}}(c) \).
\end{itemize}

The bootstrap approach provides an empirical estimate of \( c^* \) by averaging over multiple resamples, avoiding reliance on asymptotic approximations. Although computationally more intensive, it may provide robustness and flexibility for practical applications.


\section{Simulation and Numerical Illustration}
\label{sec:simulation}

The advantage of the proposed improved nonparametric estimator over the traditional nonparametric maximum likelihood estimator (MLE) is demonstrated using simulated data. We conduct a simulation study to empirically evaluate the finite-sample performance of the shrinkage-enhanced estimators in comparison to their conventional nonparametric counterparts—namely, the Product-Limit Estimator (PLE) based on system-level data and the plug-in estimator using component-level data—under the Cramer-von Mises loss function.

\subsection {Illustrative Example} \label{sec:simExample}

We begin by applying the methodology described in Section \ref{sec:improvedEstimation} to a specific coherent system. Specifically, we consider a 3-component series-parallel system, for which the structure function and component importance functions were detailed in Sections \ref{sec:egOfstructure} and \ref{sec:importance}, respectively. Component lifetimes are simulated from Weibull distributions with shape parameters \( \kappa = (2, 2, 2) \) and scale parameters \( \lambda = (2.5, 1, 1) \), corresponding to components \( k = 1, 2, 3 \). Right-censoring is introduced via a monitoring window, with censoring times \( \tau \sim \text{Exponential}(\eta = 0.05) \).

\begin{figure}[h]
\centering
\includegraphics[width=\textwidth]{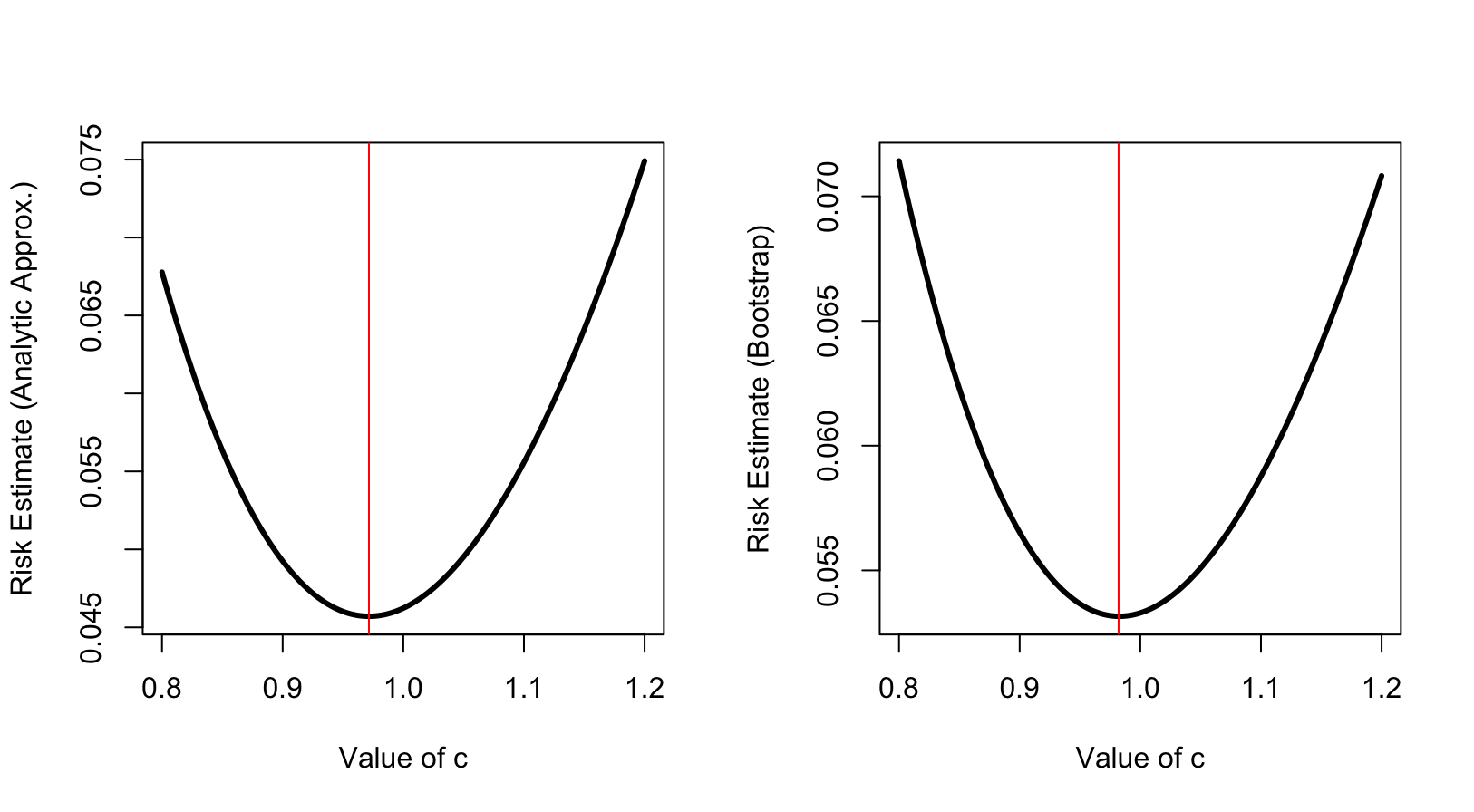}
\caption{Estimated loss for varying values of \(c\) based on a sample of 15 systems. The loss is computed using both analytical and bootstrap methods as described in Sections \ref{sec: AnalyticalC} and \ref{sec: BootstrapC}.}
\label{fig:cstar-sp3}
\end{figure}

\begin{figure}[h]
\begin{center}
\includegraphics[width=\textwidth]{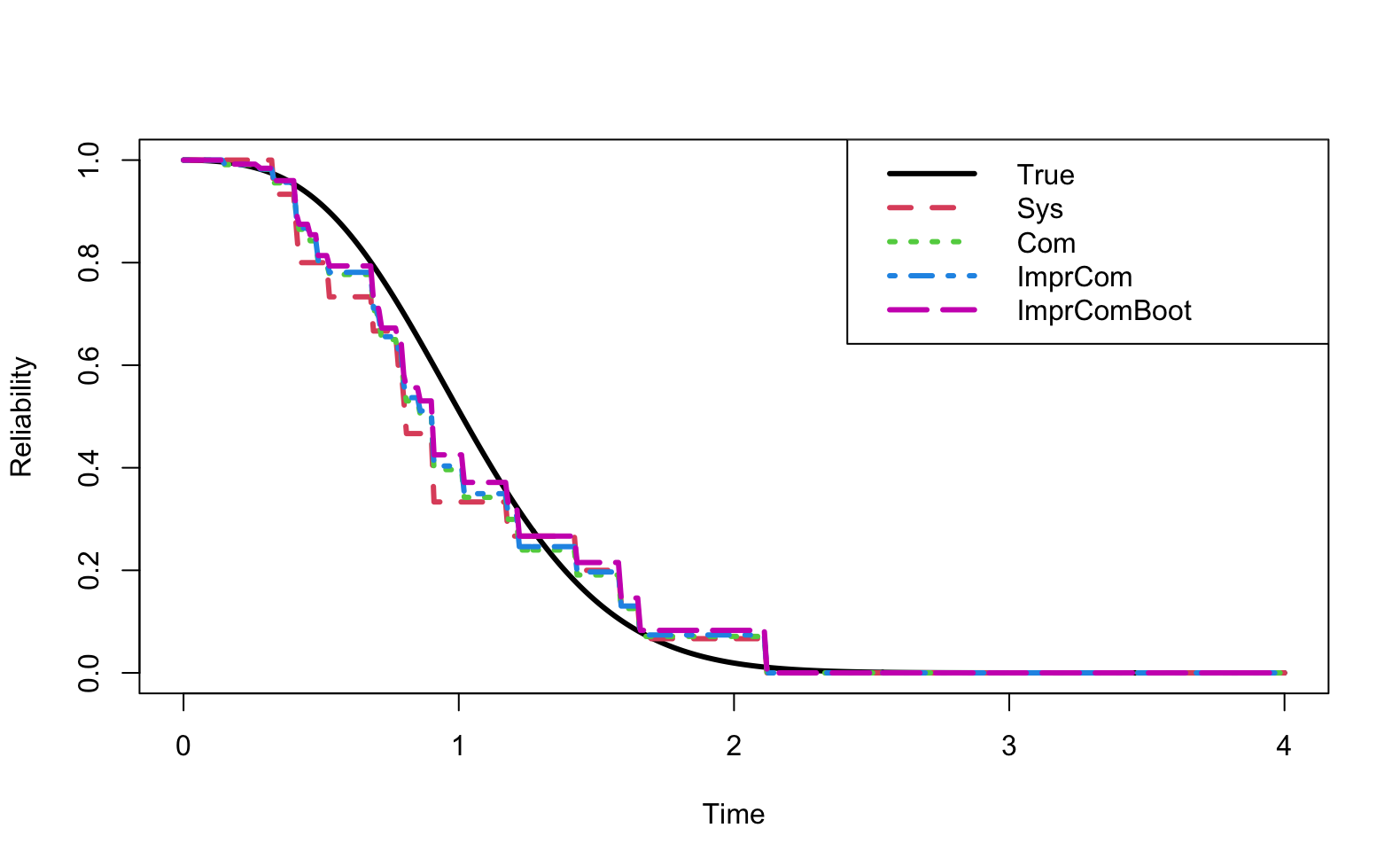}
\caption{Estimated and true system reliability functions for a 3-component series-parallel system. Estimators include the standard component-level PLE and the proposed shrinkage-enhanced estimator, based on a sample of 15 systems.}
\label{fig:est-sp3}
\end{center}
\end{figure}

Figure \ref{fig:est-sp3} shows estimated system reliability curves from one replication, alongside the true system reliability function. Figure \ref{fig:cstar-sp3} presents the estimated loss as a function of the shrinkage coefficient \( c \), computed using both analytical and bootstrap methods. The minimizing values of \( c \) are indicated.

In this replication sample, the optimal shrinkage coefficients were \( \hat{c}^* = 0.9783 \) (analytical) and \( \hat{c}^* = 0.9857 \) (bootstrap), indicating a slight upward adjustment to the component reliability estimates. While a single replication cannot provide conclusive evidence, the estimated curves demonstrate close tracking of the true reliability function, suggesting potential for improvement over conventional methods.

To assess performance more systematically, we conducted a simulation study based on 1000 Monte Carlo replications. The primary performance metric was the estimated risk, computed using the weighted Cramér–von Mises loss. We also recorded the mean and standard deviation of the estimated shrinkage coefficients \( \hat{c}^* \). Table \ref{tab:SP3} summarizes the results.

\begin{table}[h]
\centering
\caption{Mean and Standard Deviation of Estimated Risk and Shrinkage Coefficient \( \hat{c}^* \) for 3-Component Series-Parallel System (n = 15, 1000 replications).}
\label{tab:SP3}
\begin{tabular}{c|cc}
\toprule
\textbf{Method} & \textbf{Mean of \( \hat{c}^* \)} & \textbf{Estimated Risk} \\
\midrule
System PLE        & -- & 0.0713 (0.0524) \\
Component Plug-in & -- & 0.0601 (0.0465) \\
Component with Shrinkage (Analytical) & 0.984 (0.0125) & 0.0597 (0.0457) \\
Component with Shrinkage (Bootstrap) & 0.998 (0.0208) & 0.0602 (0.0461) \\
\bottomrule
\end{tabular}
\end{table}

The component-based plug-in estimator outperformed the system-level PLE, reducing the average risk by approximately 16\%. Further improvements were achieved using the analytical shrinkage method, though the gain over the standard plug-in was modest. The bootstrap-based shrinkage method performed comparably but required greater computational effort. The estimated values of \( \hat{c}^* \) remained close to 1, suggesting relatively mild shrinkage was needed in this configuration. This result reflects that the benefit of shrinkage, while present, is limited when component-level estimates are already fairly accurate and the amount of censoring is moderate. Across all methods, system-level data were nearly complete (95.07\% complete observations), whereas component-level completeness varied across components: 19.12\%, 82.97\%, and 83.06\%. This uneven censoring pattern emphasizes the importance of leveraging component-level information when it is available, particularly for more informative components.

\subsection{Simulation with Parallel and Series Systems} \label{sec:simuParaSer}

To assess generalizability across different system architectures, we expand our simulation study to examine systems with more components and structures. Specifically, we consider 5-component series and 5-component parallel systems. For each structure, we compare the following estimators:

\begin{enumerate}
    \item \textbf{System PLE}: Nonparametric MLE based solely on system-level failure data.
    \item \textbf{Component Plug-in Estimator}: Kaplan–Meier estimator (or PLE) for each component, combined via the system structure function.
    \item \textbf{Shrinkage Estimators}:
    \begin{itemize}
        \item Analytical shrinkage using the asymptotic bias-variance approximation.
        \item Bootstrap shrinkage by minimizing the empirical expected loss.
    \end{itemize}
\end{enumerate}

Each method was evaluated across 1000 simulation replications. Component lifetimes were generated from independent Weibull distributions with shape parameter \(\kappa = 2\) and scale parameter \(\lambda = 1\). Censoring was introduced via exponential monitoring times with rate \(\eta = 0.05\), resulting in moderate censoring.

Table \ref{tab:combined-risk} presents the average estimated loss (risk) and shrinkage coefficients across the replications. We also report the standard deviations in parentheses.

\begin{table}
\centering
\caption{Estimated Risk and estimated Shrinkage Coefficient \( \hat{c}^* \) for 5-Component Series and Parallel Systems (n = 15, 1000 replications). Standard deviations are provided in the brackets.}
\label{tab:combined-risk}
\begin{tabular}{lcccc}
\toprule
\textbf{System Type} & \textbf{System } & \textbf{Component } & \textbf{Shrinkage } & \textbf{Shrinkage}\\
&  \textbf{PLE}& \textbf{Plug-in} &\textbf{(Analytic)} & \textbf{(Bootstrap)}\\
\midrule
Series         & 0.0700 & 0.0700 & 0.0703& 0.0718 \\
&(0.0554) &(0.0554) & (0.0552) & (0.0564) \\
\midrule 
\(\hat c^*\) & --& -- & 0.9756 & 0.9933\\
& --&-- & (0.0029) & (0.0641) \\
\midrule
\midrule
Parallel           &  0.0714 & 0.0508& 0.0500& 0.0501 \\
&(0.0512) & (0.0420) & (0.0413) & (0.0420) \\
\midrule
\( \hat c^*\) &-- & --& 1.0141 & 1.0103 \\
&-- & --&(0.0016) & (0.0140) \\\bottomrule
\end{tabular}
\end{table}

In the series system, heavy censoring at the component level (only 19.12\% complete) meant that only one component lifetime was typically observed per system. Consequently, the component-based plug-in estimator performed identically to the system-level PLE. Shrinkage provided no meaningful improvement and, in some cases, slightly worsened performance due to over-adjustment. This highlights the limitations of shrinkage methods in data-sparse scenarios where little information is available to guide adjustment.

In contrast, the parallel system benefited from much higher component completion rates. In this setting, component-based estimators substantially outperformed the system-only PLE. Both shrinkage methods offered further gains, with the analytic method yielding the lowest average risk. The estimated shrinkage coefficients \( \hat{c}^* \) were slightly greater than 1, indicating a modest expansion in the cumulative hazard function—opposite to the slight contraction seen in the series case.

These results highlight a key insight: shrinkage effectiveness depends critically on the system structure and censoring pattern. In data-rich settings like the parallel system, shrinkage enhances estimator efficiency by fine-tuning bias–variance tradeoffs. In contrast, for highly censored or structurally constrained systems (like series systems), the benefits of shrinkage may be negligible.

\begin{figure}[h]
\begin{center}
\includegraphics[width=\textwidth]{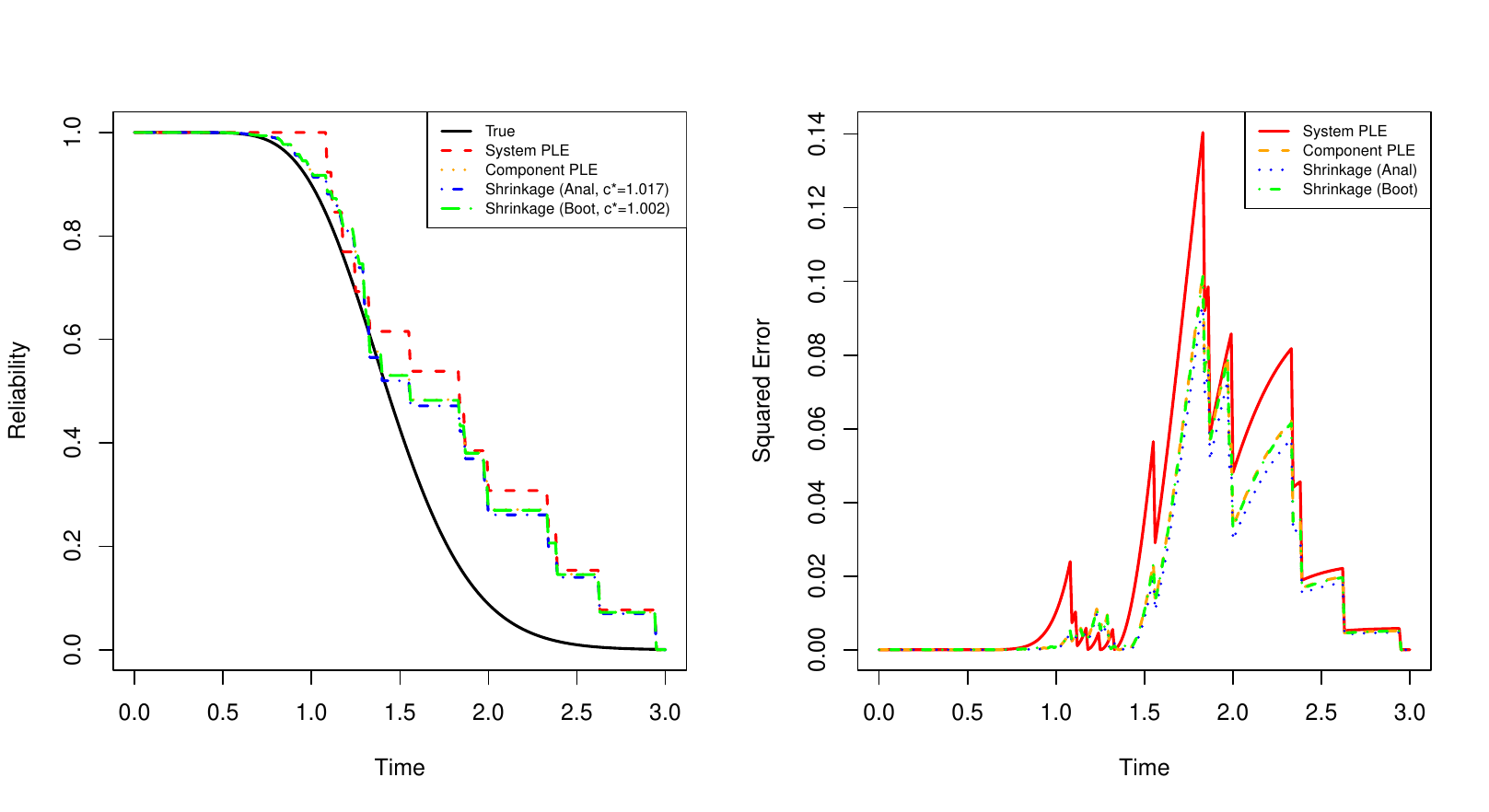}
\caption{Estimated and true system reliability functions for a 5-component parallel system. Estimators include the system-level PLE, component-based plug-in estimator, and shrinkage-enhanced estimators, based on a sample of 15 systems.}
\label{fig:NpRelEstPara}
\end{center}
\end{figure}

Figure \ref{fig:NpRelEstPara} illustrates a typical replication for the parallel system, along with the true system reliability function for a 5-component parallel system. The shrinkage-enhanced estimators track the true reliability curve more closely than the standard estimators, particularly in the mid-to-late time regions. This illustrates how shrinkage adjustments, when guided by system structure and component importance, can meaningfully improve estimation accuracy in nonparametric reliability settings.

\subsection{Effects of Sample Size, Censoring, and Number of Components}\label{sec:SimuNK}

Across all simulations, estimators that utilize component-level data consistently achieve lower risk compared to those based solely on system-level data. This reinforces the principle that detailed autopsy-level information can significantly improve the accuracy and efficiency of system reliability estimates. Moreover, when component-level estimators are enhanced via shrinkage, additional performance gains can be realized—particularly in systems with parallel components. Shrinkage techniques prove especially effective in scenarios with small sample sizes or substantial censoring. To explore these effects, we conducted simulations that varied the sample size, censoring level, and number of components \(K\), all within the framework of 5-component parallel systems, using the same baseline Weibull parameters as in earlier sections. Risk-efficiency is computed relative to the component plug-in estimator, i.e., defined as the percentage reduction in estimated risk compared to the component.

\begin{figure}[ht]
\begin{center}
\includegraphics[width=1\textwidth]{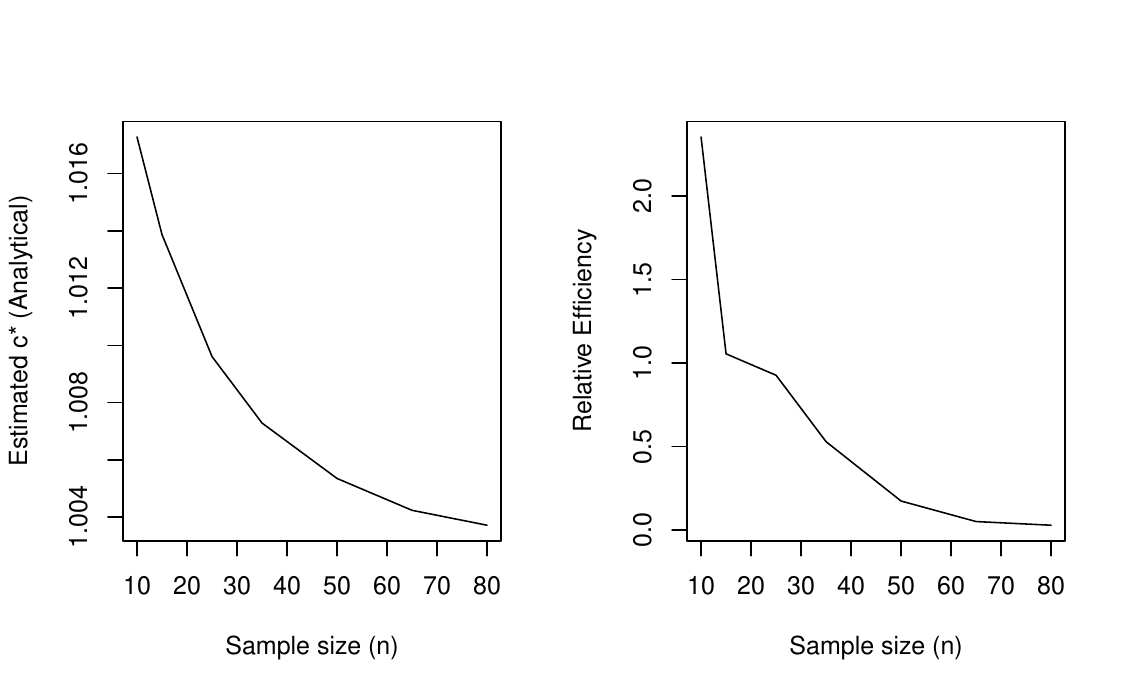}
\caption{Estimated shrinkage coefficient \( \hat{c}^* \) and relative risk-efficiency(\%) of the shrinkage estimator across varying sample sizes. Simulations assume a parallel system with 5 components and are based on 1000 replications. Risk-efficiency(\%) is computed relative to the component plug-in estimator.}
\label{fig:nEff}
\end{center}
\end{figure}

Figure \ref{fig:nEff} presents two key insights from simulations under varying sample sizes. The left panel shows the average analytically estimated shrinkage coefficient \( \hat{c}^* \), while the right panel depicts the relative risk-efficiency of the shrinkage estimator compared to the component plug-in estimator. As sample size increases, the estimated risk decreases across all estimators, reflecting improved estimation stability. Importantly, \( \hat{c}^* \) tends to converge toward 1, aligning with theoretical properties from Section \ref{sec: AnalyticalC}. In large samples, the bias inherent in nonparametric estimators diminishes, reducing the need for shrinkage or correction. Consequently, the performance advantage of shrinkage is more pronounced in small-sample regimes.

Table~\ref{tab:effectOfcensor} examines the impact of censoring, controlled via the censoring parameter \( \eta \). As \( \eta \) increases, the proportion of complete data decreases, reducing the overall information available for estimation. The shrinkage coefficient \( \hat{c}^* \) increases slightly with heavier censoring, indicating that the method compensates by effectively “inflating” the cumulative hazard estimates to account for downward bias in reliability under censoring. Interestingly, relative efficiency improves with moderate censoring levels (\( \eta = 0.1 \)), but begins to decline as censoring becomes too severe. This suggests a non-monotonic relationship: moderate levels of censoring offer an opportunity for shrinkage to improve performance, but in extremely censored settings, the benefit is limited by the sparsity of usable component information.

\begin{table}[ht]
\centering
\caption{Estimated \( \hat{c}^* \) and relative risk-efficiency(\%) of the shrinkage estimator across varying censoring levels. Simulations assume a 5-component parallel system with sample size \(n = 25\) and are based on 1000 replications. Risk-efficiency(\%) is computed relative to the component plug-in estimator.}
\label{tab:effectOfcensor}
\begin{tabular}{lccccc}
\toprule
\textbf{Censoring Parameter \( \eta \)} & \textbf{0.01} & \textbf{0.05} & \textbf{0.1} & \textbf{0.2}& \textbf{0.3} \\
\midrule
Avg. \% Complete (System) & 98.48 & 92.96 & 86.58 & 75.27 & 64.31 \\
\midrule
\( \hat{c}^* \) & 1.0096 & 1.0098 & 1.0101 & 1.0105 & 1.0108 \\
 & (0.0005) & (0.0007) & (0.0009) & (0.0011) & (0.0015) \\
\midrule
Risk-Efficiency (\%) & 0.0646 & 0.5907 & 1.1756 & 0.6638 & 0.6447 \\
\bottomrule
\end{tabular}
\end{table}

Table \ref{tab:effectOfK} examines the effect of number of parallel components on shrinkage estimation performance. As the number of components \( K \) increases from 5 to 15, we observe a gradual rise in relative efficiency gains, suggesting that the benefits of shrinkage are magnified in higher-dimensional systems. As \(K\) increases, each component's individual contribution to system reliability becomes smaller, and the aggregated estimator involves more estimated inputs. Shrinkage methods, by pooling reliability estimates across components, help stabilize the resulting system-level estimate. Interestingly, while the estimated \( \hat{c}^* \) values remain close to 1, the variance of \( \hat{c}^* \) decreases with larger \(K\), indicating that the shrinkage effect becomes more stable and predictable in higher dimensions.

\begin{table}[ht]
\centering
\caption{Estimated \( \hat{c}^* \) and relative risk-efficiency of the shrinkage estimator for parallel systems with varying numbers of components. Each simulation used \(n = 15\) systems and is based on 1000 replications. Risk-efficiency is computed relative to the component plug-in estimator.}
\label{tab:effectOfK}
\begin{tabular}{cccc}
\toprule
\textbf{Number of Components} & \( K = 5 \) & \( K = 10 \) & \( K = 15 \) \\
\midrule
\( \hat{c}^* \) & 1.0139 & 1.0140 & 1.0132 \\
 & (0.00164) & (0.00059) & (0.00041) \\
\midrule
Risk-Efficiency (\%) & 1.15 & 1.19 & 1.46 \\
\bottomrule
\end{tabular}
\end{table}

These simulations highlight some insights of shrinkage effects. Shrinkage is most beneficial under small samples and high censoring, where conventional nonparametric estimators suffer from higher variance and bias. The estimated shrinkage coefficient \( \hat{c}^* \) tends to converge to 1 as the sample size increases or censoring decreases, confirming theoretical expectations. Systems with a larger number of components tend to benefit more from shrinkage, as the aggregation of multiple noisy component-level estimates increases the potential for risk reduction via shrinkage. Taken together, these results support the use of shrinkage-enhanced nonparametric estimators as a robust and flexible tool for system reliability analysis, particularly when working with limited or heavily censored data in complex systems.

\section{Concluding Remarks}\label{sec:Conclusion}

This study introduced a new class of shrinkage estimators for nonparametric system reliability estimation, leveraging component-level information. We proposed and evaluated a novel shrinkage-enhanced methodology for estimating system reliability under right-censored data, developed within a decision-theoretic framework using the Cramer-von Mises loss function. By applying a shrinkage transformation to the component-level product-limit estimators—equivalently, shrinking the cumulative hazard estimators—we derived a more efficient estimator of the system reliability function. This approach exemplifies how simultaneous estimation can outperform individual estimation procedures when guided by a global loss criterion. Notably, while the method relies on approximations rooted in asymptotic theory, simulation results demonstrate its practical value, especially in small samples and under substantial censoring.

A key insight from this work is the critical role of data granularity in the performance of shrinkage methods. The proposed estimator showed consistent and meaningful improvements over traditional nonparametric approaches when applied to component-level (autopsy) data. This is particularly evident in systems with parallel structures, where the additional information from uncensored component failures enables a more accurate correction of bias in the plug-in estimators. Interestingly, the estimated optimal shrinkage parameter \( \hat{c}^* \) was often greater than one in such systems, suggesting that a mild “expansion” rather than shrinkage can correct underestimation induced by censoring. This observation provides a valuable perspective on how plug-in estimators can be adaptively adjusted to account for information loss.

Despite the encouraging results, several limitations remain and point toward fruitful directions for future work. First, the methodology involves reusing the observed data to estimate both the initial component reliability functions and the optimal shrinkage coefficient \( \hat{c}^* \)—a practice often referred to as “double-dipping.” While this is common in empirical Bayes and adaptive estimation frameworks, it can lead to optimistic bias in finite-sample performance evaluations. The true impact of this reuse is an ongoing area of research, particularly for small to moderate sample sizes. Nonetheless, as the sample size increases, the influence of selecting \( \hat{c}^* \) using the same data becomes asymptotically negligible. Additionally, the present study focused on a complete observation framework where component-level failure or censoring times are known. A natural and important extension is to consider current status data—a more realistic scenario in practice—where the failure status of each component is only observed at the system failure or inspection time. Developing shrinkage-enhanced estimators in this setting would require significant methodological innovations to address the limited information available on exact component failure times. 

Future research may also explore more formal Bayesian or empirical Bayes approaches, incorporating nonparametric priors such as the Dirichlet or Gamma process on the component reliability functions. These frameworks could offer an alternative to shrinkage, allowing for the inclusion of prior knowledge and yielding natural uncertainty quantification. Additionally, evaluating the performance of the proposed estimators under alternative loss functions and applying them to real-world reliability data would be important next steps for validating their broader applicability.

In this work, we provided a flexible and theoretically grounded method for improving system reliability estimation in nonparametric settings. By bridging classical shrinkage ideas with modern survival analysis, we offer a promising avenue for enhancing reliability estimation under realistic data limitations. The results underscore the utility of component-level data and the power of shrinkage strategies, especially in settings where traditional estimators may be unstable or biased.

\bibliographystyle{unsrt}
\bibliography{NpSysRel} 

\end{document}